\documentclass[reprint,twocolumn,showpacs,superscriptaddress,longbibliography,email,floatfix,aps,prl]{revtex4-2}

\usepackage{header}
\usepackage{tikz_helper}
\begin{document}

\title{Glassy dynamics with softened kinetic constraints on a noisy quantum computer}
\author{Marcel Cech}
\affiliation{Institut f\"ur Theoretische Physik and Center for Integrated Quantum Science and Technology, Universit\"at T\"ubingen, Auf der Morgenstelle 14, 72076 T\"ubingen, Germany}
\author{Igor Lesanovsky}
\affiliation{Institut f\"ur Theoretische Physik and Center for Integrated Quantum Science and Technology, Universit\"at T\"ubingen, Auf der Morgenstelle 14, 72076 T\"ubingen, Germany}
\affiliation{School of Physics and Astronomy and Centre for the Mathematics and Theoretical Physics of Quantum Non-Equilibrium Systems, The University of Nottingham, Nottingham, NG7 2RD, United Kingdom}
\author{Federico Carollo}
\affiliation{Dipartimento di Fisica, Sapienza Università di Roma, Piazzale Aldo Moro 5, 00185 Rome, Italy}

\begin{abstract}
    Mid-circuit measurements provide direct access to trajectory-level observables, revealing dynamical structures in many-body systems that are invisible in ensemble-averaged quantities. We exploit this capability to realize and study an instance of the Floquet-East model on a superconducting quantum processor. Here, the combination of mid-circuit measurements, kinetically constrained unitary operations and hardware noise gives rise to intricate many-body phenomena. Analyzing trajectories obtained from temporally and spatially resolved mid-circuit measurements, we identify dynamical heterogeneity --- a hallmark of glassy dynamics. We quantify this emergent behavior by studying the probability of finding inactive space-time regions of a given size. This quantity displays a crossover from area- to perimeter-dominated scaling, which is a characteristic property of glasses and is associated with the proximity to a dynamical first-order phase transition. Our results establish current noisy intermediate-scale quantum devices as scalable testbeds for investigating correlated many-body phenomena at the level of measurement trajectories.
\end{abstract}

\maketitle


\textbf{Introduction.---} 
Quantum devices are attracting significant interest due to their potential to simulate large quantum many-body systems that would otherwise be inaccessible with classical computers~\cite{Feynman1982,Preskill2018,Gross2017,Kim2023,Guo2024}. 
Mid-circuit measurements extend this paradigm beyond purely unitary dynamics, enabling the exploration of various facets of non-equilibrium quantum matter~\cite{Barreiro2011,Chertkov2023,Mi2024,Wu2026b,Dalmasso2026,Murauer2026}.
These operations not only influence the dynamics, but the associated outcomes---the quantum trajectories---reveal space-time resolved information about the many-body dynamics~\cite{Brighi2026,Yamamoto2026a,Yamamoto2026,Salatino2026}.
This perspective has already become a powerful tool for studying quantum-to-classical transitions~\cite{Minev2019,Guttel2026}, entanglement phase transitions~\cite{Li2023a,Garratt2024,Aziz2024,Feng2026a}, and quantum-enhanced sensing protocols~\cite{Mattes2025a,Radaelli2026}. 
The combination of mid-circuit measurements and feedback operations is furthermore a central aspect of ongoing efforts in quantum error correction~\cite{Cramer2016,RyanAnderson2021,Krinner2022}.

A particularly interesting use case for mid-circuit measurements is the study of dissipative quantum systems exhibiting glassy dynamics, reminiscent of those found in kinetically constrained models~\cite{Ritort2003,Cancrini2008,Garrahan2010a,Bodineau2012a}. 
In classical models, a hallmark of glassiness is dynamical heterogeneity, which is observable in individual stochastic realizations of the many-body dynamics. Here, different regions in space relax at vastly different rates due to a highly correlated approach towards stationarity. 
This occurs despite the fact that the stationary state itself may be entirely structureless~\cite{Jaeckle1991,Garrahan2002,Merolle2005,Jack2006,Buca2019,Katira2018,Causer2022a,Klobas2024,DeFazio2024,Sfairopoulos2025}. 
Quantum counterparts of these models exhibit slow dynamics, non-thermal eigenstates and Hilbert space fragmentation~\cite{Pancotti2020,Sala2020,ValenciaTortora2022,Brighi2023,Bertini2024}, which can be further enriched by the interplay of coherent and dissipative processes~\cite{Olmos2012,Carollo2022a,Maity2024,Marche2025,Brighi2026}. 
On classical computers, studying the ensuing quantum many-body dynamics is only possible with approximate methods, e.g., based on tensor networks ~\cite{Causer2025,Cech2025a,Cea2026,Cech2026}, that truncate the exponentially large state space. Quantum hardware, on the other hand, implements quantum dynamics and the associated state space natively. However, the current generation of quantum processors still suffers from ubiquitous noise, yielding random errors. Interestingly, these devices enable the investigation of highly nontrivial collective quantum many-body phenomena when one accepts this noise as part of the dynamics~\cite{Domingo2023,Leppaekangas2023,Guimaraes2023,Sun2024b}. 

\begin{figure}
    \centering
    \includegraphics{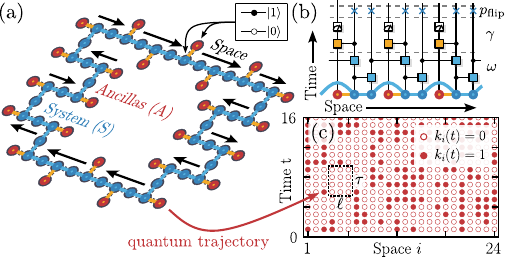}
    \caption{\textbf{Monitored circuit dynamics on the \texttt{ibm\_kingston} quantum processor.}
    (a)~Embedding of the Floquet-East model into the heavy-hex topology of the superconducting quantum processor. The system qubits (S, blue) are coupled to ancilla qubits (A, red) on every second site.
    (b)~Circuit implementation. During each discrete time step, the system first undergoes unitary evolution~[cf.~Eq.~\eqref{eq:CRX_gate}], before it is coupled to the ancillas. Blue and orange gates denote system--system and system--ancilla couplings parametrized by the quantities $\omega$ and $\gamma$, respectively. Measuring the ancillas monitors the system dynamics and results in the binary measurement outcome $k_i(t)$ describing the quantum trajectory. 
    The inherent noise of the quantum device effectively softens the kinetic constraint. 
    Its influence can be understood as the result of independent bit flips with probability $p_\mathrm{flip}$ that are indicated by the blue crosses. 
    (c)~Exemplary quantum trajectory for monitored quantum dynamics at $\omega=\pi/4$ and $\gamma=1.0$. We represent the measurement outcomes $k_i(t) = 1$ [$k_i(t) = 0$] as filled [empty] red markers.
    The rich space-time structure comprises inactive clusters that are rectangular regions of size $\ell \times \tau$ filled exclusively with `0' measurement outcomes. These clusters are the central trajectory-level observable in this work.
    }
    \label{fig:fig1}
\end{figure}

In this work, we show this exemplarily for a Floquet-East model realized on the \texttt{ibm\_kingston} quantum computer~[cf.~Fig.~\ref{fig:fig1}(a,b)]. Here, a set of ancilla qubits is coupled to a chain of system qubits that evolve under kinetically constrained dynamics.
The hardware noise predominantly gives rise to random state changes of individual qubits. 
In the context of glassy dynamics studied here, such processes effectively determine a softening of the kinetic constraint, whereby transitions that would otherwise be forbidden can occur with a small but finite probability~\cite{Shokef2010,Elmatad2010,Elmatad2013}. 
Monitoring the system dynamics takes place by periodically measuring the state of the ancillas~[cf.~Fig.~\ref{fig:fig1}(b)].
Gathering the binary measurement outcomes yields intricate space-time trajectories, an example of which is shown in Fig.~\ref{fig:fig1}(c). 
These trajectories display dynamical heterogeneity, reminiscent of glassy systems, which we characterize by analyzing the probability of finding an inactive cluster of a given spatial and temporal extent consisting exclusively of `0' measurement outcomes~[see~Fig.~\ref{fig:fig1}(c) for an example]. 
From this probability, we define a dynamical free energy and show that it exhibits a crossover from an area- to perimeter-dominated scaling as the size of the inactive cluster increases. 
This behavior is the consequence of the competition between entropic and energetic contributions near a dynamical first-order phase transition.
Our work thus highlights the utility of trajectory-level observables for investigating collective phenomena in quantum many-body systems on present-day quantum processors.


\textbf{The model.---} 
We implement the Floquet-East model on the heavy-hex topology of the \texttt{ibm\_kingston} processor as shown in Fig.~\ref{fig:fig1}(a). 
A set of $L = 48$ system~(S) qubits is arranged in a 1D spin chain with periodic boundary conditions, where each qubit is described by the computational basis $\{\ket{0}, \ket{1}\}$.
Additionally, we include $L / 2 = 24$ ancilla~(A) qubits that are adjacent to every second system qubit and which are used to monitor their dynamics~\cite{Cech2026}. 
The Floquet-East model that we consider here is then described by a series of discrete time steps, where each of them follows the quantum circuit shown in Fig.~\ref{fig:fig1}(b). 

The elementary building block is a controlled single-qubit rotation that is implemented by the unitary
\begin{align}
    u_{j-1,j} = \EastGate = e^{-i \omega n_{j-1} \sigma_j^x} = \operatorname{CRX}(2\omega)_{j-1, j} \, .
    \label{eq:CRX_gate}
\end{align}
Here, $n = \ketbra{1}$ and $\sigma^x = \ketbra{0}{1} + \mathrm{h.c.}$ denote the excitation number and the Pauli-X operators, respectively, and $\omega$ is the rotation angle.
The gate in Eq.~\eqref{eq:CRX_gate} is applied on all even and odd bonds of the spin chain in a brickwork structure. 
Notably, it only acts non-trivially on the target qubit at site $j$ if the control qubit at site $j-1$ is in state $\ket{1}$. 
This realizes the facilitation condition, which is the kinetic constraint that governs the dynamics of the East model~\cite{Jaeckle1991,Bertini2024}.
The angle $\omega$ further controls the rotation of the target qubit, with a full, i.e., deterministic, rotation occurring for $\omega = \pi/2$.

The ancilla qubits are employed to monitor the excitation number. This is achieved by preparing the ancilla qubits in state $\ket{0_\mathrm{A}}$ and coupling them to the system qubits via the same gate as in Eq.~\eqref{eq:CRX_gate}. 
With the system qubit acting as the control and the ancilla qubit as the target, a measurement of the ancilla realizes a generalized measurement on the system qubit. Formally, this can be described by a set of Kraus operators~\cite{Nielsen2010,Cech2026}. These system operators arise from projecting the joint system–ancilla evolution onto the respective ancilla measurement outcomes and are given by
\begin{equation}
    \label{eq:Kraus_operators}
    \begin{aligned}
        K_{k} &= \left(\mathds{1}_\mathrm{S} \otimes \bra{k_\mathrm{A}} \right) e^{-i \gamma n_\mathrm{S} \otimes \sigma^x_\mathrm{A}} \left(\mathds{1}_\mathrm{S} \otimes \ket{0_\mathrm{A}} \right) \\
        &= \delta_{k,0} [\mathds{1}_\mathrm{S} + (\cos \gamma - 1) n_\mathrm{S}] + \delta_{k,1} [- i (\sin \gamma) n_\mathrm{S}] \, .
    \end{aligned}
\end{equation}
Here, $\gamma$ is the angle in the system-ancilla coupling and represents the measurement strength that interpolates between projective measurements ($\gamma = \pi / 2$) and no measurements at all ($\gamma = 0$).
The measurement outcome $k_i \in \{0, 1\}$ is obtained with probability $\pi_{k_i} = \| K_{k_i} \ket{\psi} \|^2$, and the system state is updated according to $\ket{\psi} \to K_{k_i} \ket{\psi} / \| K_{k_i} \ket{\psi} \|$.
These relations emphasize that such measurements generally result in a non-linear backaction on the system, while also providing information about the ensuing dynamics in the form of the measurement record $\eta = [k_i(t)]_{i, t}$.
This quantum trajectory can exhibit a rich space-time structure even though the Floquet-East model has a structureless stationary state~(see the Supplemental Material~\cite{SM}\vphantom{\cite{JavadiAbhari2024,Krantz2019,Hines2025,Zhang2025e,Garrahan2010,Fishman2022}} for details). 
It further allows us to consider trajectory-level observables such as the activity of a given space-time region, which is defined as its total number of `1' measurement outcomes. 
In particular, trajectories of the Floquet-East model are characterized by the presence of inactive space-time clusters with zero activity~[cf.~Fig.~\ref{fig:fig1}(c)]. 

When implementing the above circuit dynamics on the \texttt{ibm\_kingston} quantum processor, the influence of the inherent hardware noise must be considered. 
Generally, noise arises from various error sources, such as imperfect gate operations, decoherence and even correlated errors, which render the characterization of corresponding noise models an ongoing challenge~\cite{Krantz2019,Hines2025,Zhang2025e}.
However, the processes that are most relevant for the Floquet-East model are random bit flips of individual qubits.
This is because they do not respect the kinetic constraint imposed by Eq.~\eqref{eq:CRX_gate}.
The ensuing dynamics on the quantum computer is therefore that of a spin model subject to softened kinetic constraints~\cite{Shokef2010,Elmatad2010,Elmatad2013}.
In the Supplemental Material~\cite{SM}, we show that our observations are consistent with an effective bit-flip probability of $p_\mathrm{flip} \approx 0.05$.

In the following, we investigate the space-time dynamics of this system, which features a total of 72 (system and ancilla) qubits. 
We analyze on the order of $10^5$ quantum trajectories that we obtained from the \texttt{ibm\_kingston} quantum processor in less than 10 minutes of runtime.
The possibility of gathering this data describing the dynamics of a quantum many-body system in a limited period of time is key to the potential of current quantum processors to investigate collective phenomena at the level of individual trajectories.


\begin{figure}
    \centering 
    \includegraphics{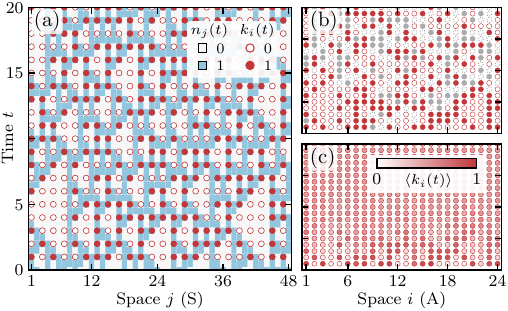}
    \caption{\textbf{Dynamics at the classical deterministic point.} (a)~Classical simulation for $\omega = \pi / 2$ and $\gamma = \pi / 2$. 
    We plot the evolution of the system in terms of local excitation numbers $n_j(t)$ at half-time steps of the gate-based dynamics.
    Empty [filled] red markers indicate the positions of projective measurements performed via the ancilla qubits, resulting in $k_i(t) = 0$ [$k_i(t) = 1$]. 
    (b)~Representative ancilla measurement outcomes obtained from a single run on the \texttt{ibm\_kingston} quantum processor for the same initial state. 
    Outcomes that match [do not match] the classical simulation are shown as red markers [light gray markers with dotted outlines]. Filled and empty markers denote $k_i(t)=1$ and $0$, respectively, as in~(a).
    (c)~Average ancilla excitation number $\expval{k_i(t)}$ of 10\,000 such trajectories. We observe that the individual measurement trajectories of the classical simulations and those recorded on the \texttt{ibm\_kingston} exhibit a pronounced dynamical heterogeneity. 
    In contrast, averaging over trajectories from the quantum processor rapidly washes out this structure because of noise-induced stochasticity. 
    }
    \label{fig:fig2}
\end{figure}

\textbf{Dynamics at the classical deterministic point.---} 
As a first step, we consider the dynamics at the special parameter set $\omega = \pi / 2$ and $\gamma = \pi / 2$, which corresponds to a classical deterministic Floquet-East model~\cite{Klobas2024}.
Here, the system state is described by a classical product state at all times, which allows for efficient classical simulations. 
Also, the ancilla measurement outcomes directly reflect the state of the corresponding system qubits.
This is because the unitary gate in Eq.~\eqref{eq:CRX_gate} is equivalent, up to a phase, to a controlled-NOT gate that implements deterministic updates between these classical states.
The Kraus operators in Eq.~\eqref{eq:Kraus_operators} instead describe projective measurements in the computational basis that extract the binary value of the corresponding system qubit without changing its state.

In Fig.~\ref{fig:fig2}(a), we show the locally resolved evolution of the excitation numbers $n_j(t)$ for a fixed, yet otherwise random, initial state in the computational basis. 
A characteristic feature of the East model is the formation of triangular void regions with $n_j(t) = 0$. 
These regions are a manifestation of the dynamical heterogeneity: the kinetic constraint allows for these regions to only dissolve starting from their left boundary, which renders them long-lived. 
Regions with an extensive number of excitations relax much faster.
As ancilla measurement outcomes $k_i(t)$ are obtained directly from the excitation number $n_j(t)$ at every second qubit, they display a very similar space-time structure~[cf.~red markers in Fig.~\ref{fig:fig2}(a)].
In particular, inactive clusters of exclusively `0' measurement outcomes are clearly visible and mark space-time regions with different dynamical behavior.

Running the same circuit on the \texttt{ibm\_kingston}, we obtain measurement trajectories such as the one shown in Fig.~\ref{fig:fig2}(b). 
We see that this measurement record exhibits individual defects~(see the light-gray ancilla outcomes) that arise from the hardware noise and make the dynamics on the quantum processor non-deterministic.
This is also clearly seen in the average over 10\,000 trajectories that we show in Fig.~\ref{fig:fig2}(c). 
Although deterministic and noise-free dynamics would produce an average identical to the ancilla outcomes shown in panel~(a), here, the average ancilla excitation number $\expval{k_i(t)}$ rapidly approaches a uniform value that provides no information about potential collective phenomena.
However, individual trajectories still display a highly correlated space-time structure that represents the dynamical heterogeneity. 
This is also the case away from this special point~[cf.~Fig.~\ref{fig:fig1}(c)], where, in addition to the stochasticity arising from hardware noise, we have quantum projection noise due to the fact that the system state is no longer represented by a classical product state.


\begin{figure}
    \centering 
    \includegraphics{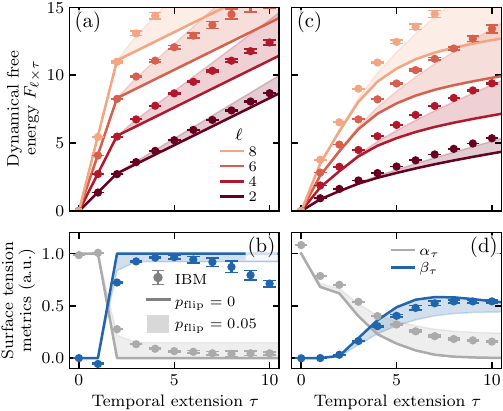}
    \caption{\textbf{Statistics of inactive clusters in space-time.} (a)~Negative log-probability $F_{\ell \times \tau}$ of finding an inactive space-time cluster of size $\ell \times \tau$ at the classical deterministic point ($\omega = \pi / 2$ and $\gamma = \pi / 2$). For fixed spatial extensions $\ell$~(see color reported in the legend), we plot the dynamical free energy against the temporal extension $\tau$. We compare results obtained from 200\,000 quantum trajectories that we recorded on the \texttt{ibm\_kingston} quantum processor (markers with error bars indicating twice the standard error of the mean) with tensor-network calculations of perfect and softened kinetic constraints~(lines and shaded regions with $p_\mathrm{flip} = 0$ and $p_\mathrm{flip} = 0.05$, respectively). 
    (b)~Corresponding surface tension metrics. We plot the values of $\alpha_\tau$ and $\beta_\tau$ that we extract from the fits according to Eq.~\eqref{eq:surface_tension_inactive_cluster} over $\ell \in \{2, ..., 5\}$ in gray and blue. 
    These quantities capture the area- and perimeter-related contributions to the dynamical free energy $F_{\ell \times \tau}$. 
    (c,d)~Same as in the panels~(a,b), but for monitored quantum dynamics at $\omega = \pi / 4$ and $\gamma = 1.0$. 
    The increase of $\beta_\tau$ over $\alpha_\tau$ at late times indicates the crossover to a perimeter-dominated scaling, which is a hallmark of dynamical heterogeneity.
    Note that the surface tension metrics are normalized according to independent measurement outcomes of the same average activity. For details, see the Supplemental Material~\cite{SM}.
    }
    \label{fig:fig3}
\end{figure}

\textbf{Statistics of inactive clusters in space-time.---} 
To analyze collective behavior, we specifically focus on the statistics of inactive clusters, which are defined as rectangular regions of size $\ell \times \tau$ in the measurement record $\eta$ that contain exclusively `0' measurement outcomes~[cf.~Fig.~\ref{fig:fig1}(c)].
At the trajectory level, their probability $p_{\ell \times \tau}$ is a multi-point correlation function. 
In an uncorrelated trajectory, the probability of finding an inactive cluster of size $\ell \times \tau$ factorizes and hence decays exponentially with the area $\ell \tau$ of the cluster.
We can draw a thermodynamic analogy, where we interpret the binary measurement record $\eta$ as a microscopic configuration of a 1+1D spin system~\cite{Garrahan2010}.
This allows us to define the dynamical free energy of an inactive cluster as $F_{\ell \times \tau} = -\log p_{\ell \times \tau}$. 
Here, the exponential decay of the probability with the area of the cluster can be understood as an entropic free-energy cost associated with the formation of this ordered structure, which is proportional to its area $\ell \tau$.
Indeed, as we show in Fig.~\ref{fig:fig3}(a,c) for the classical deterministic point and the monitored quantum dynamics, the dynamical free energy for clusters with small temporal extent displays a slope that is proportional to their spatial extension~$\ell$. 
This means that the dynamical free energy scales as the area of these clusters.
However, as the temporal extension $\tau$ of the cluster increases, we observe that the dynamical free energy $F_{\ell \times \tau}$ grows more slowly, and its dependence on the spatial extension $\ell$ becomes progressively weaker.
Hence, the probability is no longer suppressed exponentially with the area of the cluster, but rather with its perimeter.
This crossover in the scaling behavior of inactive clusters is due to the fact that trajectories look random on a small scale, but are highly correlated on a mesoscopic scale. 

The phenomenology is reminiscent of the so-called hydrophobic~\cite{Chandler2005,Chandler2010}, or generally orderphobic~\cite{Katira2016,Lunn2026}, effect in the vicinity of a first-order phase transition.
In this analogy, inactive clusters play the role of solutes or ordered regions. 
Small inactive clusters are formed by random fluctuations and are therefore described by an entropic free-energy cost. 
In contrast, large inactive clusters are stabilized by the reduction in free-energy cost associated with creating an interface between the ordered and disordered phases.
For quantum trajectories that display unexpectedly large inactive clusters, this means that the system is dynamically close to a first-order phase transition between an active and an inactive phase in trajectory space~\cite{Katira2018}. 
At the phase transition, macroscopically large active and inactive space-time regions coexist and give rise to the dynamical heterogeneity, which is a hallmark of glassy dynamics.
Here, the appearance of large inactive clusters merely comes at the free-energy cost associated with the interface to the active phase, which determines the perimeter-law scaling of the dynamical free energy $F_{\ell \times \tau}$.
For the classical and the monitored quantum Floquet-East model with perfect kinetic constraints, this dynamical first-order phase transition is known to exist at the level of trajectories~\cite{Klobas2024,DeFazio2024,Cech2026}. 

In our setup, we have a softening of the East-model constraint due to the inherent noise on the \texttt{ibm\_kingston} quantum processor, which is expected to move the system away from the exact phase coexistence point~\cite{Elmatad2010,Elmatad2013}. This leads to an intricate balance between entropic and energetic contributions to the dynamical free energy $F_{\ell \times \tau}$.
Specifically, the formation of large inactive clusters in these circumstances involves stochastic contributions arising from potential spin flips, as well as interfacial stabilization connected to kinetically constrained dynamics.
Concretely, we have
\begin{align}
    \label{eq:surface_tension_inactive_cluster}
    \Delta_\tau F_{\ell \times \tau} = F_{\ell \times (\tau + 1)} - F_{\ell \times \tau} = \alpha_\tau \ell + \beta_\tau \, ,
\end{align}
where $\alpha_\tau$ and $\beta_\tau$ are fit parameters encoding the area- and perimeter-related contributions to the dynamical free energy, respectively. 
Figure~\ref{fig:fig3}(b,d) shows the extracted values of $\alpha_\tau$ and $\beta_\tau$ for the classical deterministic point and the monitored quantum dynamics, respectively.
We observe that the crossover from area- to perimeter-dominated scaling occurs at $\tau = 2$ for the classical deterministic point~(see the Supplemental Material~\cite{SM} for the analytical derivation of the noise-free case).
For the monitored quantum dynamics, this crossover is more gradual, but we still find that $\beta_\tau$ eventually exceeds $\alpha_\tau$, which indicates the emergence of a perimeter-dominated scaling reminiscent of strong spatiotemporal correlations at a mesoscopic scale.
Together with the observation that the area contribution, as quantified by $\alpha_\tau$, is non-zero, this indicates that the system is indeed not exactly at the phase coexistence point but is sufficiently close to witness the competition between the two scalings. 
These observations are also in line with the tensor-network-based calculations of the dynamical free energy for a softened kinetic constraint with $p_\mathrm{flip} = 0.05$ displayed in Fig.~\ref{fig:fig3}~(see also the Supplemental Material~\cite{SM} for details).
This supports the interpretation that the dominant effect of noise on these trajectory-level observables can be understood in terms of independent bit flips, thereby emphasizing that the quantum processor is effectively simulating softened kinetic constraints.


\textbf{Discussion and outlook.---} In this work, we have demonstrated that mid-circuit measurements on a noisy quantum processor provide access to trajectory-level dynamical heterogeneity in a kinetically constrained quantum system. The observed crossover behavior in the statistics of inactive clusters reveals pronounced spatiotemporal correlations that signal a nearby first-order dynamical phase transition. Hardware noise is an ingredient of the overall dynamics, and its effect is a softening of the kinetic constraint.

To the best of our knowledge, our work is the first study of quantum many-body systems on a noisy quantum processor in which stochastic outcomes of mid-circuit measurements are subject to a direct statistical many-body analysis. This represents a complementary direction to their established application in characterizing quantum states, as in studies of entanglement phase transitions, or in controlling quantum states through feedback and error correction protocols.
It opens the door to investigate increasingly complex quantum many-body dynamics at the level of their individual quantum trajectories. Such explorations will require a careful understanding of how hardware noise affects the dynamics and the measurement records, while also raising the possibility of harnessing noise as a dynamical resource. As this progresses, recent data-driven approaches, such as those proposed in Refs.~\cite{Brunner2025,Fitzner2026}, may complement handcrafted observables by identifying hidden order parameters and unexpected features in the trajectories of intricate quantum dynamics.


\textbf{Code and Data availability.---} 
The code and data supporting the findings of this work are available on Zenodo \cite{ZenodoData}.


\let\oldaddcontentsline\addcontentsline 
\renewcommand{\addcontentsline}[3]{}

\acknowledgments
\textbf{Acknowledgements.---} 
We thank Cecilia De Fazio, María Cea and Mari Carmen Ba\~{n}uls for fruitful discussions. F.C. is grateful to Filippo Gambetta for valuable discussions. We acknowledge the use of the \texttt{ibm\_kingston} quantum processor (one of IBM's publicly accessible Heron r2 superconducting transmon devices with 156 qubits in a heavy-hex connectivity) under the open plan agreement. The views expressed in this work are those of the authors and do not reflect the official policy or position of IBM or the IBM Quantum Team. Numerical simulations were performed using the ITensor library~\cite{Fishman2022}. We acknowledge funding from the Deutsche Forschungsgemeinschaft (DFG, German Research Foundation) through the Research Unit FOR 5413/1, Grant No. 465199066, and through the Research Unit FOR 5522/1, Grant No. 499180199. We also acknowledge support from the Leverhulme Trust (Grant No. RPG-2024-112). This work is supported by ERC grant OPEN-2QS (Grant No. 101164443, https://doi.org/10.3030/101164443).

\textbf{AI usage.---} 
The authors acknowledge using GitHub Copilot (powered by Anthropic's Claude Sonnet 4.6) and OpenAI's ChatGPT 5.5 to make minor improvements to the presentation of figures and for language editing. All suggestions were reviewed and approved by the authors.


\bibliography{biblio.bib}

\let\addcontentsline\oldaddcontentsline 




\clearpage
\onecolumngrid

\setcounter{equation}{0}
\setcounter{page}{1}

\setcounter{figure}{0}
\setcounter{table}{0}
\setcounter{section}{0}
\makeatletter
\renewcommand{\theequation}{S\arabic{equation}}
\renewcommand{\thefigure}{S\arabic{figure}}
\renewcommand{\thetable}{S\arabic{table}}

\begin{center}
{\Large SUPPLEMENTAL MATERIAL}
\end{center}
\begin{center}
\vspace{0.8cm}
{\Large Glassy dynamics with softened kinetic constraints on a noisy quantum computer}
\end{center}
\begin{center}
Marcel Cech,$^{1}$ Igor Lesanovsky,$^{1,2}$ and Federico Carollo$^{3}$ 
\end{center}
\begin{center}
$^1${\em Institut f\"ur Theoretische Physik and Center for Integrated Quantum Science and Technology, }\\
{\em Universit\"at T\"ubingen, Auf der Morgenstelle 14, 72076 T\"ubingen, Germany}\\
$^2${\em School of Physics and Astronomy and Centre for the Mathematics}\\
{\em and Theoretical Physics of Quantum Non-Equilibrium Systems,}\\ 
{\em The University of Nottingham, Nottingham, NG7 2RD, United Kingdom}\\
$^3${\em Dipartimento di Fisica, Sapienza Università di Roma, Piazzale Aldo Moro 5, 00185 Rome, Italy}
\end{center}

\tableofcontents

\section{Properties of the monitored quantum Floquet-East model}
In this section, we provide further details on the properties of the Floquet-East model and the implementation on the \texttt{ibm\_kingston} quantum processor. 
First, we discuss the properties of the ensemble-averaged state and demonstrate that the fully mixed state is a stationary state of the dynamics. Next, we elaborate on the concrete implementation of the circuit dynamics before providing details on the bit-flip noise channel as an effective description of the softened kinetic constraint on the quantum processor.

\subsection{Average state dynamics}
We start by discussing the properties of the ensemble-averaged state and show that the fully mixed state is a stationary state of the dynamics, independently of the specific parameters. We argue that this allows us to leverage Hadamard-like rotations paired with projective measurements to sample the statistics of inactive clusters at stationarity. \\

The average state dynamics can be described in three steps that follow the circuit representation shown in Fig.~\ref{fig:fig1}(b). First, by averaging over all ancilla measurement outcomes, the combination of unitary dynamics and monitoring via the ancilla qubits is described by
\begin{align}
    \label{eq:quantum_channel}
    \rho \to \mathcal{E}_{\omega, \gamma}[\rho] = \sum_{\{\mathbf{k}\}} K_{\mathbf{k}} U \rho U^\dagger K_{\mathbf{k}}^\dagger \, .
\end{align}
Here, $U$ is the unitary evolution constructed from the Floquet-East model gates in Eq.~\eqref{eq:CRX_gate}, and $K_\mathbf{k}$ abbreviates the product of the single-site Kraus operators in Eq.~\eqref{eq:Kraus_operators} corresponding to the measurement outcomes $\mathbf{k} = [k_i]_i$.

To model the dominant influence of the hardware noise, we consider independent and incoherent bit flips~(see Sec.~\ref{sec:noise_model} for a more detailed discussion). 
For bit flips that occur with a probability of $p_\mathrm{flip}$ per system qubit and time step, this is described by bit-flip quantum channels according to
\begin{align}
    \label{eq:bit_flip_channel}
    \rho \to \mathcal{E}_{p_\mathrm{flip}}[\rho] = \mathcal{E}_{p_\mathrm{flip}}^{(1)} \circ ... \circ \mathcal{E}_{p_\mathrm{flip}}^{(L)} [\rho] \, .
\end{align}
Here, $\mathcal{E}_{p_\mathrm{flip}}^{(j)}[\rho] = (1 - p_\mathrm{flip}) \rho + p_\mathrm{flip} \sigma^x_j \rho \sigma^x_j$ is the bit-flip quantum channel of site $j$.

Since the Kraus operators in Eq.~\eqref{eq:Kraus_operators}~[and also those associated with Eq.~\eqref{eq:bit_flip_channel}] are normal operators, it is easy to see that these quantum channels are unital, i.e., they map the fully mixed state $\mathds{1} / 2^L$ to itself. Hence, the fully mixed state is a stationary state of the combined channel $\mathcal{E}_{p_\mathrm{flip}} \circ \mathcal{E}_{\omega, \gamma}$ for any values of $\omega$, $\gamma$, and $p_\mathrm{flip}$. \\

This property allows us to sample trajectory-level observables such as the probability of inactive clusters at stationarity by initializing the system qubits via the combination of Hadamard-like rotations and measurements in the computational basis.

It is important to note that, when $p_\mathrm{flip} = 0$, the fully mixed state is not the only stationary state. The perfect kinetic constraint combined with periodic boundary conditions dynamically decouples the completely de-excited state~$\ket{0}^{\otimes L}$, which gives rise to an exceptional stationary state~\cite{Marche2025}. However, its weight according to the fully mixed state is exponentially suppressed in system size, and we never observed it in the data that we obtained on the \texttt{ibm\_kingston}.
As it further ceases to be a stationary state when $p_\mathrm{flip} > 0$, we neglect its influence and only comment on it again when discussing the analytical solution of the dynamical free energy of the classical deterministic Floquet-East model at $p_\mathrm{flip} = 0$.

\subsection{Implementation of the circuit dynamics on the \texttt{ibm\_kingston} quantum processor}
In this section, we discuss the implementation of monitored circuit dynamics on the \texttt{ibm\_kingston} quantum processor. 
Generally, we follow the quantum circuit outlined in Fig.~\ref{fig:fig1}(b) that only features gates between neighboring qubits according to the mapping in Fig.~\ref{fig:fig1}(a). \\

For the concrete implementation, we make two additional considerations.
First, if no initial state is fixed manually, we can study the stationary distribution of trajectory-level observables by exploiting the sampling procedure for the initial state outlined above.  
Importantly, we only need to compile a single circuit that can be run over and over again. 
Second, we notice that the monitoring described in Eq.~\eqref{eq:Kraus_operators} is invariant under initializing an ancilla in state $\ket{1}$ and swapping the labels $k_i(t) \in \{0, 1\}$ of the ancilla measurement outcomes. 
This is because the interaction between system and ancilla qubit can be written as 
\begin{align}
    \begin{aligned}
        U_{\mathrm{SA}} &= e^{-i \gamma n_\mathrm{S} \otimes \sigma^x_\mathrm{A}} \\
        &= \left[\mathds{1}_\mathrm{S} + (\cos \gamma - 1) n_\mathrm{S} \right] \otimes \mathds{1}_\mathrm{A} - i [\sin \gamma] n_\mathrm{S} \otimes \sigma^x_\mathrm{A} \, .
    \end{aligned}
\end{align}
Here, the first part leaves the ancilla qubit in the same state as before the interaction, while the second part flips the ancilla qubit. 
Therefore, and even though it would be technically possible, we eliminate the need to reset the ancillas by labeling a change in the ancilla state as a `1' measurement outcome and no change as a `0'. 
We note that this is indeed similar to the common definition of activity in classical kinetically constrained models, where these state changes are among the central quantities of interest. \\

The resulting circuits were then compiled using the Qiskit library~\cite{JavadiAbhari2024} and submitted to the \texttt{ibm\_kingston} quantum processor on June 18, 2026 within the 10 free runtime minutes of the open plan agreement. 
Here, the circuits for 40 time steps at $(\omega, \gamma) = (\pi / 2, \pi / 2)$ and $(\omega, \gamma) = (\pi / 4, 1.0)$ have circuit depths of approximately 360 and 1040, respectively.
Note further that we use 10\,000 shots for Fig.~\ref{fig:fig2}, while extending this number to 200\,000 per parameter set for the analysis of the dynamical free energy in Fig.~\ref{fig:fig3}.

\subsection{Bit-flip noise channel and softened kinetic constraints}
\label{sec:noise_model}
We now review the effective noise model of individual and uncorrelated bit flips that we outlined in the main text. 
We also elaborate on the specific value of $p_\mathrm{flip} = 0.05$, which we selected for the comparison in Fig.~\ref{fig:fig3}.\\

First, we clarify again that it is not the purpose of this noise model to describe the exact noise of the quantum processor but rather to aid the understanding of the ensuing dynamics. 
The development of accurate noise models is an ongoing challenge due to various noise mechanisms including slightly imperfect pulse sequences, spurious couplings between qubits on the device or between qubits and their environment~\cite{Krantz2019,Hines2025,Zhang2025e}.
However, in order to understand the Floquet-East model as implemented on the \texttt{ibm\_kingston} quantum processor, it is most important to consider the effective processes that can violate the kinetic constraints encoded in Eq.~\eqref{eq:CRX_gate}.
These are the processes that do not require an excited qubit to change the state of its neighbor~\cite{Shokef2010,Elmatad2010,Elmatad2013}.
This class includes, for example, incoherent independent bit flips that are parametrized by a single probability $p_\mathrm{flip}$ according to Eq.~\eqref{eq:bit_flip_channel}. \\

\begin{figure}[ht]
    \centering
    \includegraphics{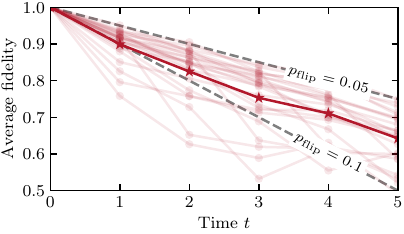}
    \caption{\textbf{Error analysis of the ancilla measurement outcomes for the classical deterministic dynamics.} We plot the average ancilla fidelity as obtained by comparing the ancilla measurement outcomes $k_i(t)$ obtained on the \texttt{ibm\_kingston} quantum processor to the ideal ancilla measurement outcomes for the classical deterministic Floquet-East model with $\omega = \pi / 2$ and $\gamma = \pi / 2$. 
    Based on 5 different initial conditions with 10\,000 trajectories each, we show the average over all ancilla qubits (red line with star markers) and the average over individual ancilla qubits (faint red lines with circle markers). The dashed black lines show the expected decrease in fidelity from random, independent bit flips with probabilities of $p_\mathrm{flip} = 0.05$ and $p_\mathrm{flip} = 0.1$ per system qubit and time step without considering the propagation of errors or more than one bit flip per site.}
    \label{fig:fig2_SM_error_analysis}
\end{figure}

In Fig.~\ref{fig:fig2_SM_error_analysis}, we now consider the fidelity of ancilla measurement outcomes at the classical deterministic point. 
This entails comparing $k_i(t)$ that we obtained on the quantum processor to the classically calculated $k_i^\mathrm{cl}(t)$. The average fidelity is then computed as $1 - \expval{|k_i^\mathrm{cl}(t) - k_i(t)|}$, where the average runs over the number of trajectories as well as potentially the number of qubits. 
In Fig.~\ref{fig:fig2_SM_error_analysis}, we observe that this quantity initially decreases approximately linearly with time.
This is consistent with the fact that, for short times and to first order in the bit-flip probability $p_\mathrm{flip}$, the ancilla fidelity is sensitive to bit flips in the system but insensitive to the interplay of multiple of them.
As such, the slope is set by the bit-flip probability $p_\mathrm{flip}$.
Focusing on the slope of the ancilla fidelity averaged over all ancillas, we observe that the decrease is consistent with an effective bit-flip probability of $p_\mathrm{flip} \approx 0.05$. 
The steeper decrease at $t = 1$ can be attributed to additional factors in state preparation and readout that, if continued, would overestimate the later decrease.
Note further that also in Fig.~\ref{fig:fig3}, the predictions of the associated dynamics are in line with the observations made for the statistics of inactive clusters.

\subsection{Parameter choice for the presented dynamics}
Finally, we discuss our choice of $\omega$ and $\gamma$ in the main text. 
On the one hand, $\omega = \pi / 2$ and $\gamma = \pi / 2$ correspond to the classical deterministic Floquet-East model, and permit the direct comparison of the classically calculated ancilla measurement outcomes and those obtained on the \texttt{ibm\_kingston} quantum processor as done in Fig.~\ref{fig:fig2} upon choosing the same initial state. Furthermore, the probability of inactive regions can be calculated analytically~(see, in particular, Sec.~\ref{sec:analytic_inactive_cluster_det_classical}), where we find that the dynamical free energy under these perfect and deterministic kinetic constraints exhibits an instantaneous area-to-perimeter scaling crossover at $\tau = 2$.

On the other hand, we choose $\omega = \pi / 4$ and $\gamma = 1.0$ for the monitored quantum dynamics that we study in Fig.~\ref{fig:fig1}(c) and Fig.~\ref{fig:fig3}(c,d).
This choice is motivated by two considerations. First, the measurement strength specified by $\gamma \neq \pi / 2$ allows for coherent superpositions in the system dynamics due to the non-projective nature of $K_{0}$ in Eq.~\eqref{eq:Kraus_operators}.
Second, choosing $\omega$ sufficiently far from the deterministic point makes the dynamics markedly stochastic.
Combined with $\gamma = 1.0$, it remains in a regime in which exploratory numerical simulations have suggested that the crossover from area- to perimeter-scaling in the temporal extension of inactive clusters occurs within an accessible time frame.

\section{Trajectories, activity and dynamical phase transitions}
In this section, we provide additional details on the importance of ancilla measurement outcomes as quantum trajectories in the analysis of monitored quantum dynamics. We further discuss the role of inactive space-time clusters as trajectory-level observables and how their statistics reflect dynamical first-order phase transitions linked to kinetic glass formers. 

\subsection{The dynamical activity phase diagram}
As a starting point, we define the activity and describe its role as an order parameter for dynamical phase transitions. Note that the discussion is adapted from Ref.~\cite{Cech2026} and Ref.~\cite{Klobas2024} for the monitored quantum and the classical deterministic Floquet-East model, respectively.

\begin{figure}[ht]
    \centering
    \includegraphics{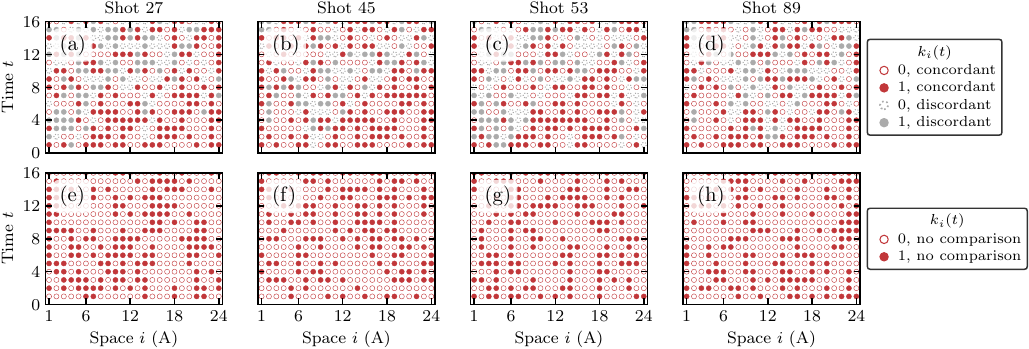}
    \caption{\textbf{Exemplary ancilla measurement outcomes from the \texttt{ibm\_kingston}.} (a-d)~Ancilla measurement outcomes $k_i(t)$ for the classical deterministic Floquet-East model with $\omega = \pi / 2$ and $\gamma = \pi / 2$. (e-h)~Same as in panel~(a-d), but for the monitored quantum dynamics with $\omega = \pi / 4$ and $\gamma = 1.0$. 
    Note that the initial state for all panels is the same as in Fig.~\ref{fig:fig2}. 
    For panels~(a-d), this allows for a direct comparison to the measurement outcomes shown in Fig.~\ref{fig:fig2}(a). We thereby highlight those that match or do not match the ones from classical simulations in red or light gray, respectively. In panels~(e-h), the initial state is identical, however, quantum projection noise is expected to produce different trajectories in different realizations. Therefore, no comparison is made.
    }
    \label{fig:figSX_L48_single_shots}
\end{figure}

As set out in the main text, the central objects in this work are quantum trajectories that we obtain by the ancilla measurement records $\eta = [k_i(t)]_{i,t}$ recorded on the \texttt{ibm\_kingston} quantum processor. 
First, mid-circuit measurement capabilities and high repetition rates on this device make it possible to access these objects in large numbers~[see Fig.~\ref{fig:figSX_L48_single_shots} for a few examples].
Furthermore, they are free of postselection overheads and allow for an unambiguous definition of activity in terms of the total number of `1' measurement outcomes in a given space-time region. 
Considering this quantity over the full extent of a quantum trajectory $\eta$, the time-integrated activity yields
\begin{align}
    A_{L,T}[\eta] = \sum_{\{i,t\}} k_i(t) \, ,
\end{align}
where the indices $i$ and $t$ run over the $L/2$ ancillas and $T$ simulated time steps, respectively.
Note that we include the system size $L$ and the total number of time steps $T$ to later distinguish it from the thermodynamic limit with $L, T \to \infty$.
Crucially, employing a thermodynamic analogy between trajectories and microcanonical configurations~\cite{Garrahan2010}, we define the dynamical partition function
\begin{align}
    \label{eq:partition_function}
    \mathcal{Z}_{L,T}(s) = \sum_{\{ \eta \}} \pi[\eta] e^{-s A_{L,T}[\eta]} \, ,
\end{align}
where $\pi[\eta]$ is the probability of a measurement record $\eta$ and $s$ is a counting field that biases the trajectories according to their activity. 
If not stated otherwise, this probability is defined in terms of an initial state that is sampled from a stationary state. 

By definition, the dynamical partition function is the moment generating function of the activity. 
In terms of the large deviation formalism, it is useful to introduce the scaled cumulant generating function $\theta_{L, T}(s)$ as $\mathcal{Z}_{L, T}(s) = e^{\frac{LT}{2} \theta_{L, T}(s)}$.
A dynamical phase transition is encoded in the scaled cumulant-generating function $\theta_{L, T}(s)$ developing non-analytic behavior as one approaches the thermodynamic limit $L, T \to \infty$.
As discussed in Ref.~\cite{Cech2026}, when every system qubit is monitored, $\theta(s) = \lim_{L, T \to \infty} \theta_{L, T}(s)$ exhibits a discontinuity in the first derivative at $s = 0$ for both classical and monitored quantum dynamics in this Floquet-East model.
This marks a first-order phase transition in the trajectory space and manifests in the dynamical phase coexistence of the active and inactive phases in individual trajectories. \\

Although monitoring is implemented only on every second qubit, we expect that the monitored quantum dynamics under perfect kinetic constraints exhibits a similar dynamical first-order phase transition at $s = 0$ in the thermodynamic limit.
This is supported by two observations that contribute to the arguments for the dynamical phase transition discussed in Ref.~\cite{Klobas2024}:
First, the average activity density is finite, and when the initial state is sampled from the fully mixed state, this value is simply given by $-\theta_{L, T}'(0) = 2 \expval{A_{L, T}[\eta]}_\eta / LT = (\sin^2\gamma) / 2$ for all $L$ and $T$.
This establishes a finite slope of the scaled cumulant generating function at $s = 0$.
Second, as seen in Fig.~\ref{fig:fig3}(b,d) for $p_\mathrm{flip}=0$, the area-contribution to the dynamical free energy labeled by $\alpha_\tau$ vanishes as $\tau$ increases. 
Assuming that this scaling extends to larger and larger inactive clusters, the corresponding trajectories eventually dominate the dynamical partition function in Eq.~\eqref{eq:partition_function} for $s > 0$ and thus enforce $\theta(s>0) = 0$. 
However, with the finite slope at $s = 0$ mentioned above, this leads to a discontinuity of the first derivative of $\theta(s)$.
We note that this reasoning, together with the interpretation in terms of the hydrophobic effect, underlines the relation of the scaling-crossover and its significance in signaling emergent effects in quantum trajectories.

\subsection{Statistics of inactive clusters}
With this background, we now discuss the statistics of inactive clusters in more detail.
After deriving the analytical expression for the probability of an inactive cluster at the classical deterministic point, we provide a more detailed discussion of the evaluations leading to Fig.~\ref{fig:fig3} of the main text. 

\subsubsection{Probability of inactive clusters at the classical deterministic point}
\label{sec:analytic_inactive_cluster_det_classical}
Following the ideas established in Ref.~\cite{Klobas2024}, we now derive the analytical expression for the probability of an inactive cluster at the classical deterministic point, i.e., for $\omega = \pi / 2$ and $\gamma = \pi / 2$. In particular, we show that the probability of an inactive cluster of size $\ell \times \tau$ is given by
\begin{align}
    \label{eq:prob_inactive_cluster_deterministic_endmatter}
    p_{\ell \times \tau}^\mathrm{det} = \begin{cases}
        2^{- \ell \tau} \, , & \text{if $\tau < 2$} \\
        2^{- 2(\ell - 1) - \tau} \, , & \text{if $\tau \geq 2$}
    \end{cases}
\end{align}
where we can further extract the surface tension coefficients
\begin{align}
    \label{eq:surface_tension_metrics_inactive_cluster_deterministic_endmatter}
    \alpha_\tau^\mathrm{det} = \begin{cases}
        \log{2} \, ,  \\
        0 \, , 
    \end{cases}\&
    \quad
    \beta_\tau^\mathrm{det} = \begin{cases}
        0 \, , & \text{if $\tau < 2$} \\
        \log{2} \, , & \text{if $\tau \geq 2$}
    \end{cases} \, .
\end{align}
The significance of this result is that it shows that the crossover from area- to perimeter-scaling occurs at $\tau = 2$ for the classical deterministic Floquet-East model. 

For the sake of completeness, we note that for finite $L$, the probability of an inactive cluster saturates at $2^{-L}$, which corresponds to the dynamically decoupled state $\ket{0}^{\otimes L}$. 
However, the suppression with system size renders this contribution negligible for the system sizes considered in this work, and we can perform the following considerations without explicitly considering it. \\

As a starting point for the calculations, we recall that the classical deterministic Floquet-East model is defined by the parameters $\omega = \pi / 2$ and $\gamma = \pi / 2$. Here, the dynamics of an initial state in the computational basis is deterministically described in terms of the controlled-NOT gate, which we represent as follows
\begin{align}
    \label{eq:deterministic_East_gate_endmatter} 
    u_{j-1, j} \to \EastGateDeterministic \equiv \operatorname{CNOT}_{j-1, j}\, .
\end{align}
Furthermore, we introduce the following symbols to describe the monitoring. 
In particular, due to the choice of $\gamma = \pi / 2$, this extracts the local excitation number at the corresponding site according to
\begin{align}
    \label{eq:deterministic_measurement_endmatter}
    \begin{aligned}
        K_{i, 0} &\to \ProjectiveMeasurementCircleZero \equiv \ketbra{0} \, ,\\
        K_{i, 1} &\to \ProjectiveMeasurementCircleOne \equiv \ketbra{1} \, .
    \end{aligned}
\end{align}

\begin{figure}
    \centering
    \includegraphics{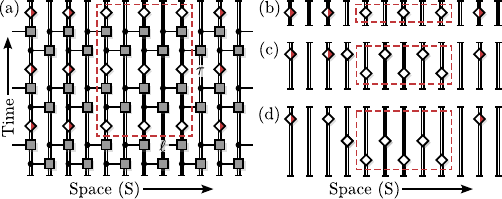}
    \caption{\textbf{Probability of inactive clusters at the classical deterministic point.} (a)~Exemplary diagrammatic representation of the probability $p_{\ell \times \tau}^\mathrm{det}$ of finding an $\ell \times \tau = 3 \times 3$-sized inactive cluster in the ancilla measurement record for the classical deterministic Floquet-East model. The red dashed rectangle indicates the $\ell \times \tau$-sized inactive cluster, where all ancilla outcomes are conditioned to output zero. (b-d)~Contraction of the diagram in panel~(a) for $\tau = 1, 2, 3$. Note that we have not absorbed unconditioned ancilla outcomes into the trace, as a guide to the eye. }
    \label{fig:figSX_p_det}
\end{figure}

To then calculate the probability of an inactive cluster, it is convenient to consider the folded space, i.e., the space of the density matrix, which allows us to represent the dynamics in a diagram as shown in Fig.~\ref{fig:figSX_p_det}(a)~\cite{Klobas2024,Cech2026}. 
Introducing the symbols for the local fully mixed state $\mathds{1} / 2 \equiv \InfiniteTemperatureLine$ and the partial trace of a qubit $\Tr{\cdot} \equiv \TraceLine$, the probability of an inactive cluster is given by the value of the contracted diagram specified by the corresponding spatial and temporal extension of $\ell$ and $\tau$.
Note that this represents the trace over the time-evolved stationary state under dynamics that are either conditioned to output zero~(cf.~$\MeasurementDiamondZeroDoubleLine$), if inside the $\ell \times \tau$-sized inactive cluster, or unconditioned~(cf.~$\MeasurementDiamondDoubleLine$) otherwise.
For the classical deterministic Floquet-East model, this calculation can be done analytically by applying the following diagrammatic relations:
\begin{align}
    \label{eq:diagramatic_rules}
    \begin{aligned}
        (\mathrm{I}_\mathrm{a})~&\UnitalRuleCNOT\, , &\quad (\mathrm{I}_\mathrm{b})~&\UnitalRuleMeasurement \, ,\\
        (\mathrm{I}_\mathrm{c})~&\UnitalRuleCNOTUpsideDown\, , &\quad (\mathrm{I}_\mathrm{d})~&\UnitalRuleMeasurementUpsideDown \, ,\\
        (\mathrm{II}_\mathrm{a})~&\ZeroControlCNOT\, , &\quad (\mathrm{II}_\mathrm{b})~&\ZeroControlTarget \,,\\
        (\mathrm{III}_\mathrm{a})~&\ProjectorOnMixedState\, , &\quad (\mathrm{III}_\mathrm{b})~&\ContractionMixedState \, .\\
    \end{aligned}
\end{align}
In particular, these relations can be understood as follows: $(\mathrm{I}_\mathrm{a})$ and $(\mathrm{I}_\mathrm{b})$ state that the evolution under the controlled-NOT gate and the projective measurement are unital, i.e., they map the fully mixed state onto itself. This can be used to evaluate the diagram from the bottom to its top. 
Similarly, but due to trace preservation, we have the relations $(\mathrm{I}_\mathrm{c})$ and $(\mathrm{I}_\mathrm{d})$, which can be used to contract the diagram from top to bottom. 
Furthermore, $(\mathrm{II}_\mathrm{a})$ states that zeros on the control qubit of a controlled-NOT gate factorize the dynamics without affecting the state of the other qubit, while $(\mathrm{II}_\mathrm{b})$ states that zeros on the target qubit before and after the controlled-NOT gate require that the control qubit is also in state zero. This is a direct consequence of the strict facilitation condition encoded in the controlled-NOT gate.
Finally, the relations $(\mathrm{III}_\mathrm{a})$ and $(\mathrm{III}_\mathrm{b})$ describe the contraction of the diagram, which amounts to taking the trace over the time-evolved stationary state, which then counts the fraction of the remaining states.

With this toolbox of diagrammatic relations in Eq.~\eqref{eq:diagramatic_rules}, we can now contract the diagram in Fig.~\ref{fig:figSX_p_det}(a) for arbitrary $\ell$ and $\tau$. 
For $\tau = 1$, we immediately notice that the first layers of controlled-NOT gates can be contracted using $(\mathrm{I}_\mathrm{a})$. This amounts to Fig.~\ref{fig:figSX_p_det}(b) and hence~[using the relations $(\mathrm{II}_\mathrm{a})$ and $(\mathrm{II}_\mathrm{b})$], we obtain $p_{\ell \times 1}^\mathrm{det} = 2^{-\ell}$.
Continuing to $\tau = 2$, we first use $(\mathrm{III}_\mathrm{a})$ to disconnect the cluster from everything to its right, after which we can simplify the diagram wherever there are no projections to $\ket{0}$. 
Hence, using $(\mathrm{III}_\mathrm{b})$ this enforces that all $2\ell - 1$ qubits within the red dashed rectangle and the first qubit to its left are in state $\ket{0}$. 
The corresponding diagram is shown in Fig.~\ref{fig:figSX_p_det}(c) and we therefore obtain $p_{\ell \times 2}^\mathrm{det} = 2^{-2\ell}$.
Given that now all qubits within the red dashed rectangle and one qubit to its left in Fig.~\ref{fig:figSX_p_det}(c) are projected to zero, $\tau = 3$ requires that also the second qubit to the left is projected to zero, which yields the diagram in Fig.~\ref{fig:figSX_p_det}(d) and hence $p_{\ell \times 3}^\mathrm{det} = 2^{-2\ell - 1}$. Iterating this procedure while using $(\mathrm{III}_\mathrm{b})$~[and potentially $(\mathrm{I}_\mathrm{c,d})$ for intermediate contractions] then increases the number of conditioned system qubits by one for each further time step. Hence, we find that for $\tau \geq 2$, the probability of an inactive cluster is given by $p_{\ell \times \tau}^\mathrm{det} = 2^{-2(\ell - 1) - \tau}$ as described in Eq.~\eqref{eq:prob_inactive_cluster_deterministic_endmatter}.

\subsubsection{Details on the evaluation of inactive clusters from quantum trajectories}
Next, we turn to the evaluation of the dynamical free energy $F_{\ell \times \tau} = -\log p_{\ell \times \tau}$ of inactive clusters from the ancilla measurement records $\eta = [k_i(t)]_{i,t}$ obtained on the \texttt{ibm\_kingston} quantum processor.

Generally, we evaluate the probability of an inactive cluster of size $\ell \times \tau$ by counting the number of occurrences of these clusters in the ancilla measurement records $\eta = [k_i(t)]_{i,t}$ and dividing it by the total number of possible clusters of size $\ell \times \tau$ in the trajectories. 
In particular, this amounts to a space-time average over the entire trajectory, i.e., over all $L / 2$ ancilla measurement outcomes with periodic boundaries and all 40 simulated discrete time steps. \\

\begin{figure}[ht]
    \centering
    \includegraphics{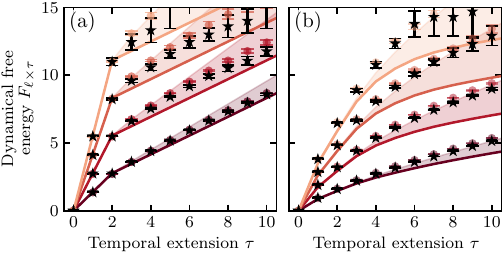}
    \caption{\textbf{Influence of space-time average in the statistics of inactive clusters in space-time.} (a) Dynamical free energy $F_{\ell \times \tau}$ of inactive space-time clusters of size $\ell \times \tau$ for the classical deterministic point ($\omega = \pi / 2$ and $\gamma = \pi / 2$). As in Fig.~\ref{fig:fig3}(a), we compare results obtained from 200\,000 trajectories with 40 time steps (markers with error bars indicating twice the standard error of the mean, colors indicate $\ell = 2, 4, 6, 8$, respectively) with theoretical calculations (solid lines and shaded area). In addition, we show the extracted value of $F_{\ell \times \tau}$ if we consider the same number of trajectories but only the first $\tau$ time steps (dark gray stars). Panel~(b) shows the same analysis for the monitored quantum dynamics at $\omega = \pi / 4$ and $\gamma = 1.0$. }
    \label{fig:fig3_SM_comparison_traj_clipped}
\end{figure}

To assess the influence of this space-time average on the dynamical free energy, we turn to Fig.~\ref{fig:fig3_SM_comparison_traj_clipped}.
Here, we compare the results presented in Fig.~\ref{fig:fig3}(a,c) with the dynamical free energy $F_{\ell \times \tau}$ that we infer from only the first $\tau$ time steps.
Comparing the two, we find that this space-time average does not significantly influence the results for the statistics of inactive clusters. 
This holds true for both the classical deterministic point and the monitored quantum dynamics shown in Figs.~\ref{fig:fig3_SM_comparison_traj_clipped}(a) and~(b), respectively.
It demonstrates that the inactive clusters are a stable feature of the dynamics and that the space-time average can be employed to reduce the statistical error in the empirical analysis of the quantum trajectories.
If not stated otherwise, we therefore evaluate quantum trajectories over their whole space-time extent.
\\

\begin{figure}[ht]
    \centering
    \includegraphics{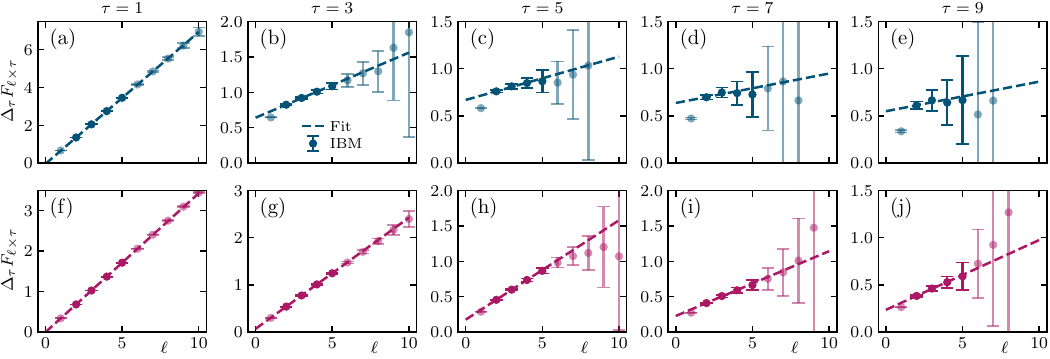}
    \caption{\textbf{Surface tension metrics.} (a-e) Temporal surface tension $\Delta_\tau F_{\ell \times \tau} = F_{\ell \times (\tau+1)} - F_{\ell \times \tau}$ as a function of $\ell$ for different values of $\tau$ for the classical deterministic Floquet-East model at $\omega = \pi / 2$ and $\gamma = \pi / 2$. We compare the results obtained from the trajectories on the \texttt{ibm\_kingston} quantum processor (markers with error bars indicating twice the standard error of the mean) to the extracted linear fit (dashed lines) that allows us to extract the surface tension metrics $\alpha_\tau$ and $\beta_\tau$ according to Eq.~\eqref{eq:surface_tension_inactive_cluster}. Note that only the opaque markers for $\ell \in \{2, ..., 5\}$ are used for the linear fit, while the semi-transparent markers are shown for additional information. Panels (f-j) show the same analysis for the monitored quantum dynamics at $\omega = \pi / 4$ and $\gamma = 1.0$.}
    \label{fig:fig3_SM_fit_quality}
\end{figure}

Using the extracted dynamical free energy, we consider the temporal surface tension $\Delta_\tau F_{\ell \times \tau} = F_{\ell \times (\tau + 1)} - F_{\ell \times \tau}$~[cf.~Eq.~\eqref{eq:surface_tension_inactive_cluster}].
We show this quantity in Fig.~\ref{fig:fig3_SM_fit_quality} to illustrate the fit procedure with which we determine the surface tension metrics shown in Fig.~\ref{fig:fig3}(b,d).
We observe that the temporal surface tensions for inactive clusters are in agreement with the ansatz in Eq.~\eqref{eq:surface_tension_inactive_cluster}.
As such, $\alpha_\tau$ and $\beta_\tau$ correspond to the slope and y-intercept of the dashed line, respectively. 
Note that $\ell = 1$ is deliberately excluded from the fit due to the fact that this primarily probes a single point in space, rather than an extended region like the others. 

Concerning the slightly negative value of $\beta_\tau$ for $\tau=1$ in Fig.~\ref{fig:fig3}(b), we attribute this to the corresponding fitting procedure shown in Fig.~\ref{fig:fig3_SM_fit_quality}(a). This is because the combination of a steep increase in the temporal surface tension and a practically vanishing offset makes it likely that slightly negative, yet physically irrelevant, values of $\beta_\tau$ arise. \\

To improve readability in Fig.~\ref{fig:fig3}(b,d), we show the surface tension metrics $\alpha_\tau$ and $\beta_\tau$ in arbitrary units, which we obtain by dividing the extracted fit parameters by $(-\log(1 - (\sin^2\gamma) / 2))$.
This normalization is based on the value of $\alpha_\tau$ for random and uncorrelated measurement outcomes of the same average activity of $(\sin^2\gamma) / 2$, where the probability of an inactive cluster is given by $p_{\ell \times \tau} = (1 - (\sin^2\gamma) / 2)^{\ell\tau}$.

\subsubsection{Details on tensor network simulations}
Finally, we discuss the tensor network simulations that we use for the comparison in Fig.~\ref{fig:fig3}.
The calculation entails contracting the diagram in Fig.~\ref{fig:figSX_p_det}(a) when generalized to the values of $\omega$ and $\gamma$ that are considered.
In particular, we follow the approach outlined in Ref.~\cite{Cech2026}. Using the ITensor library~\cite{Fishman2022}, we encode the fully mixed state as a matrix product operator before time-evolving it using the time-evolving block decimation (TEBD) algorithm according to the above-mentioned description. 
Additionally, we can incorporate the bit-flip channel in Eq.~\eqref{eq:bit_flip_channel} with a given $p_\mathrm{flip}$ if desired.
The precision of the simulation is controlled by the maximally allowed bond dimension~$\chi$. For $\chi = 64$, we found good convergence for all the results shown in Fig.~\ref{fig:fig3}.

\end{document}